\documentclass[manuscript, screen]{acmart}

\usepackage{verbatim}       
\usepackage{xcolor}         
\usepackage{hyperref}       
\usepackage{url}            
\usepackage{booktabs}       
\usepackage{amsfonts}       
\usepackage{enumitem}

\usepackage{mathtools}

\usepackage{subcaption}

\usepackage{graphicx}
\usepackage{tabularx}
\usepackage{adjustbox}
\usepackage{longtable}

\usepackage{algorithm}
\usepackage[noend]{algpseudocode}

\usepackage{comment}

\def\1{\bm{1}}

\newcommand{\etal}{\textit{et~al.}\@}

\AtBeginDocument{%
  }

\copyrightyear{2026}
\acmYear{2026}
\setcopyright{cc}
\setcctype{by}
\acmConference[OzCHI '26]{Proceedings of the 38th Australasian Conference on Human-Computer Interaction}{November 21--25, 2026}{Adelaide, SA, Australia}
\acmBooktitle{Proceedings of the 38th Australasian Conference on Human-Computer Interaction (OzCHI '26), November 21--25, 2026, Adelaide, SA, Australia}
\acmDOI{10.1145/3847401.3857149}
\acmISBN{979-8-4007-2724-5/2026/11}

\begin{document}

\title{CaRing: Preventing Carpal Tunnel Syndrome based on Daily Activities from Always-Available Input Device}

\author{Shuowei Li}
\email{lis26@uw.edu}
\orcid{0000-0003-4934-4691}
\affiliation{%
  \institution{University of Washington}
  \city{Seattle}
  \state{WA}
  \country{USA}
}

\author{Houdong Liang}
\affiliation{%
  \institution{University of California, San Diego}
  \city{La Jolla}
  \state{California}
  \country{USA}
  }
\email{holiang@ucsd.edu}
\orcid{0000-0003-2438-745X}

\author{Xingjian Dong}
\affiliation{%
  \institution{University of Southern California}
  \city{Los Angeles}
  \country{USA}}
\email{xdong404@usc.edu}
\orcid{0009-0007-9527-9575}

\renewcommand{\shortauthors}{Li et al.}

\begin{abstract}
We present CaRing, a ring worn on the base knuckle of the index finger, a wearable system for detecting the start and end of mouse use to help prevent Carpal Tunnel Syndrome, in which the damage to the median nerve is permanent. CaRing senses finger movement, which neither a software timer nor a wrist-worn device detects. The displacement reported by an optical flow sensor is accumulated into a running value, then a zero point is measured while the hand rests on the desk at the start of each session. With this formulation, the start and end thresholds are expressed relative to the session's zero point. CaRing does not introduce any per-user parameter. We empirically demonstrate that approximately $90\%$ of start and end events are detected within two seconds of the researcher's label, using 35 recordings and a lab study with ten users.
\end{abstract}

\ccsdesc[500]{Human-centered computing~Ubiquitous and mobile devices}
\ccsdesc[300]{Human-centered computing~Pointing devices}
\ccsdesc[100]{Human-centered computing~Empirical studies in ubiquitous and mobile computing}

\keywords{Carpal Tunnel Syndrome, Repetitive Strain Injury, Wearable Computing, Always-Available Sensing, Finger-Worn Sensor, Optical Flow Sensor, Mouse Use Detection, Occupational Health}


\maketitle

\section{Introduction}
Carpal Tunnel Syndrome (CTS) is a compressive neuropathy of the median nerve at the wrist~\cite{tosti2012acute}. A key symptom of CTS is intermittent numbness of all fingers except the little finger, and the damage to the median nerve is permanent once it develops. To reduce the pressure on the median nerve, users are encouraged to take breaks between tasks~\cite{mayoclinic_carpaltunnel_2026} and to maintain good posture~\cite{cole2006reducing}. More than $48\%$ of computer users inadvertently lift their index and middle fingers while using a mouse~\cite{lee2008observed}, a habit that brings pain in the hand and forearm.

Preventing CTS during daily computer use presents two challenges. First, the harm follows from how the hand moves, and not only from how long a session runs. Software timers and ambient displays prompt a break after a set interval~\cite{mclean2001microbreaks, mateevitsi2014healthbar}, and RSIGuard monitors keyboard and mouse activity~\cite{rsiguard}. These systems observe only the events the computer produces and never see the fingers. Second, the sensors that observe the hand are attached to the desk or to the wrist. Camera-based systems measure posture and hand activity~\cite{jaimes2005sit, radwin2023} but only while the hands remain in view, an instrumented mouse carries the sensor in the equipment rather than on the user~\cite{tobias2013exploring, bhaskar2004ergonomic}, and a wrist-worn device measures the wrist angle and stops above the knuckles~\cite{leung2012limber}. Neither kind of system records the finger movements that accompany nerve compression.

We developed CaRing, a ring worn on the base knuckle of the index finger toward always-available sensing of hand use. In this work, CaRing detects when the mouse is in use but does not yet assess posture or movement quality.
 Starting from the raw output of one optical flow sensor, CaRing calibrates a zero point at the beginning of each session. A threshold on the y-axis signal then marks the start and end of each period of mouse use, and the system warns the user when a period reaches 15 minutes. Because the sensor sits on the finger, CaRing needs no specially designed mouse and no camera in front of the workspace, and the same ring follows the user between the keyboard and the mouse. CaRing also records every period of use, giving medical researchers a usage log that self-reports cannot provide. We derived the design parameters from 35 recordings totaling about 3.2 hours of sensor data, and evaluated CaRing in a separate 30-minute lab study with 10 computer users. 
In the lab study, CaRing detected approximately $90\%$ of the start and end events within two seconds of the researcher's labels. In a post-session questionnaire, participants estimated an average of nine minutes of mouse use, while CaRing recorded thirteen minutes. The participants underestimated their own mouse use by $1.4\times$ ($p = .006$). Therefore, for CTS prevention, it is important to sense mouse use directly rather than rely on the user's own estimate. To the best of the authors' knowledge, CaRing is the first CTS prevention system to detect mouse use by sensing finger movement.
 
\begin{figure}[!htbp]
    \centering
    
    \begin{subfigure}[b]{0.4\textwidth}
        \centering
        \includegraphics[width=\textwidth]{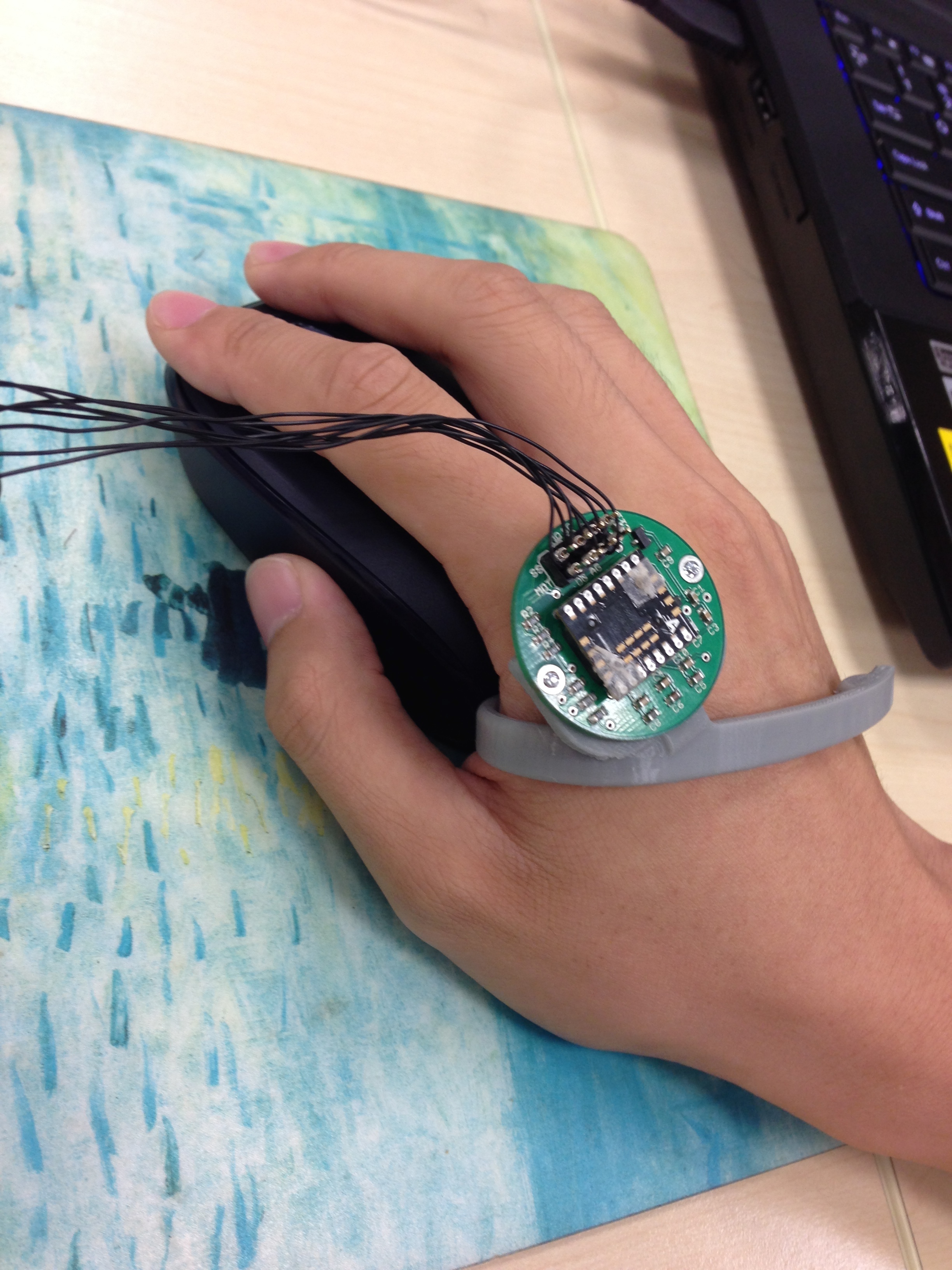}
        \Description{Mouse usage.}
        \caption{Mouse usage. The CaRing prototype senses the optical flow signal generated by finger movement while a user operates a mouse.}
        \label{subfig:wear_ring}
    \end{subfigure}
    \hfill 
    \begin{subfigure}[b]{0.4\textwidth}
        \centering
        \includegraphics[width=\textwidth]{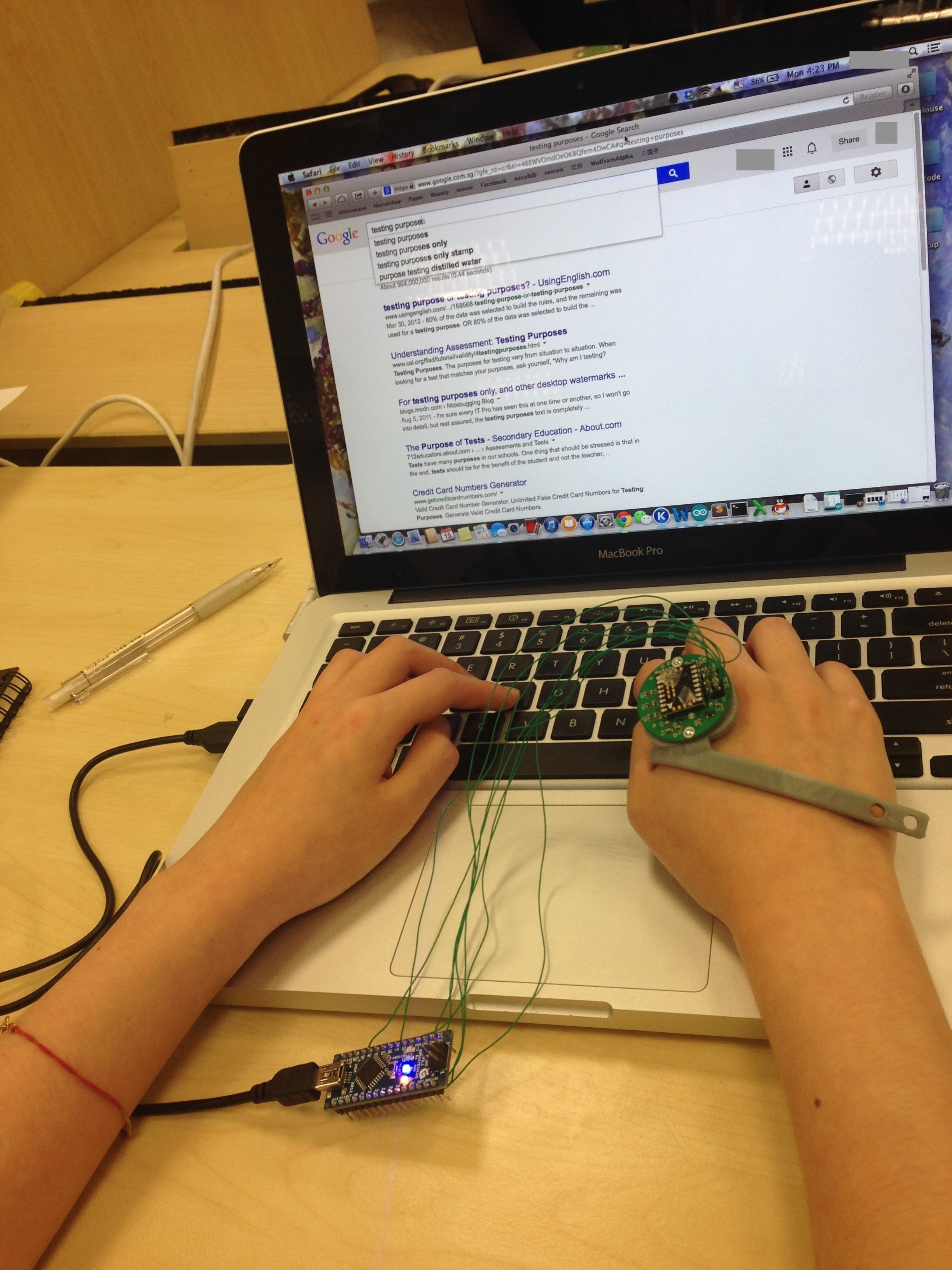}
        \Description{Keyboard typing.}
        \caption{Keyboard typing. The CaRing prototype senses the optical-flow signal produced by finger movement while a user types on a computer keyboard.}
        \label{subfig:typing}
    \end{subfigure}
    
    \caption{CaRing Prototype. The wearable system uses an optical flow sensor mounted on the base knuckle of the index finger to sense finger-movement signals during daily computer tasks.}
    \label{fig:main_figure}
\end{figure}

\section{Related Work}
\label{sec:related}
\subsection{Software Sensing}
\label{subsec:software_sense}
Software sensing infers a user's state from signals the computer already produces during ordinary use, such as keystrokes, mouse clicks, and elapsed time, without any dedicated sensor. By tracking how long users spend on computers, such software can remind them to take proper breaks.

McLean \etal~\cite{mclean2001microbreaks} compared microbreak schedules during computer terminal work, and reported less discomfort when short breaks are inserted at regular intervals. Mateevitsi \etal~\cite{mateevitsi2014healthbar} used an ambient display that shows accumulated sedentary time as a health bar. The display changes how the record is presented, but the record still comes from elapsed time. Commercial products such as RSIGuard~\cite{rsiguard} monitor computer activity and suggest break times. However, the monitoring is based on computer use alone and does not account for the hand postures that cause CTS.

Such background signals form long time series. Time series analysis remains an active area of research. ~\cite{ni2024timeseries} studied heart rate forecasting. Several existing works identify abnormal data, including outlying samples under information constraints~\cite{hu2025infocons}, anomalous event traces~\cite{xu2026chainawareencodingmicroservicetrace}, and harmful covariate shifts~\cite{ginsberg2022learning}. Han et al. proposed a method to decide how long to observe before acting~\cite{han2026earlyearlyenoughdesigndependent}.

\subsection{Environmental Sensing}
\label{subsec:environmental_sense}
A second approach places a physical sensor in one fixed location, either a camera facing the workspace or a piece of equipment. Thus, the user's hand must come to the sensor.

The posture alarm system by Jaimes~\citep{jaimes2005sit} places a camera in front of the workspace, provides real-time posture feedback, and raises an alarm when posture is poor, but only works when the user faces the camera. 
Related work counts repetitive actions from body motion~\cite{gu2025mocount} and decodes actions with motion-language~\cite{JIA2026115077} and vision-language-action models~\cite{zhang2026liralocalcrosslayerinformation}. However, a sensor outside the body must still observe the hand.
Radwin \etal~\cite{radwin2023} use markerless computer vision to measure hand speed and exertion frequency, and to compute the Hand Activity Level, a common CTS risk metric. Across $419$ industrial videos, the measure closely tracks traditional single-frame analysis. Such a pipeline could label training data for a ring through cross-modal distillation~\cite{liang2026oc}, with active~\cite{yuan2022opticalflow, zeng2026does} or semi-supervised learning~\cite{zhou2024reducing} reducing labeling costs. However, the hands must still stay in view.

A sensor can also be embedded in equipment. The habit-aware mouse~\cite{tobias2013exploring}, a spongy mouse, and an armrest mouse pad~\cite{bhaskar2004ergonomic} each add sensing to desk equipment. Thus, the benefit accrues to the equipment rather than to the user, who needs it both at home and in the office.
Watanabe \etal~\cite{watanabe2021} take a similar approach in clinical practice, where a support vector machine separates CTS and non-CTS cases from stylus trajectory and pressure recorded while a user draws a spiral on a tablet, as machine learning does in other clinical screening tasks~\cite{chang2025pdunet}. The method relies on a structured drawing task rather than everyday use, so it is a screening test, not a background monitor.

\subsection{Wearable Sensing}
\label{subsec:wearable_sense}
A third approach moves the sensor onto the body, so it travels with the user instead of staying at the desk. Limber~\citep{leung2012limber} combines a torso-mounted enclosure, a hoodie, and a wristband to detect body posture and wrist angle and to warn the user on the computer when a harmful habit is detected; however, users found it uncomfortable to wear.

Moschetti \etal~\cite{moschetti2016} take a lighter approach. Finger- and wrist-worn inertial measurement units (IMUs) recognize daily gestures without any deliberate action, so the recognition can flag repetitive actions during ordinary use. The authors report how accuracy varies with sensor placement on the index finger, thumb, and wrist.
Mennella \etal~\cite{mennella2022} evaluate how arrays of physical sensors, including IMUs and pressure sensors, are combined for hand functional assessment and catalogs metrics that could set kinetic thresholds for a preventive device.

Outside the sensing literature, cognitive models reproduce skilled hand movements, showing that they are structured and repeatable~\cite{he2025abacus}.

Still, existing wearables stop short of the finger. The Limber wristband measures the wrist, and the IMUs of Moschetti \etal are evaluated for gesture recognition rather than for monitoring mouse use.

Across all three approaches, no existing system keeps a sensor on the finger throughout a session. More specifically, software sensing has no physical signal at all, environmental sensing observes the hand only near a camera or other equipment, and wearable sensing stops at the wrist or torso, above the knuckles. CaRing closes this gap by placing its one sensor directly on the finger, following the user between keyboard and mouse and sensing how the hand moves, not just how long it has been used.
 
\section{System Implementation Details}
\label{sec:system}
The CaRing prototype captures raw data from the ADNS-9800 optical flow sensor embedded in the ring. Then transmits it to a host computer, which detects the start and end of each period of mouse use. After 15 minutes of continuous use, the host program displays a warning message reminding the user to adjust their posture or take a break.

\subsection{Hardware Implementation}
\begin{figure}[!htbp]
    \centering
    
    \begin{subfigure}[b]{0.4\textwidth}
        \centering
        \includegraphics[width=\textwidth]{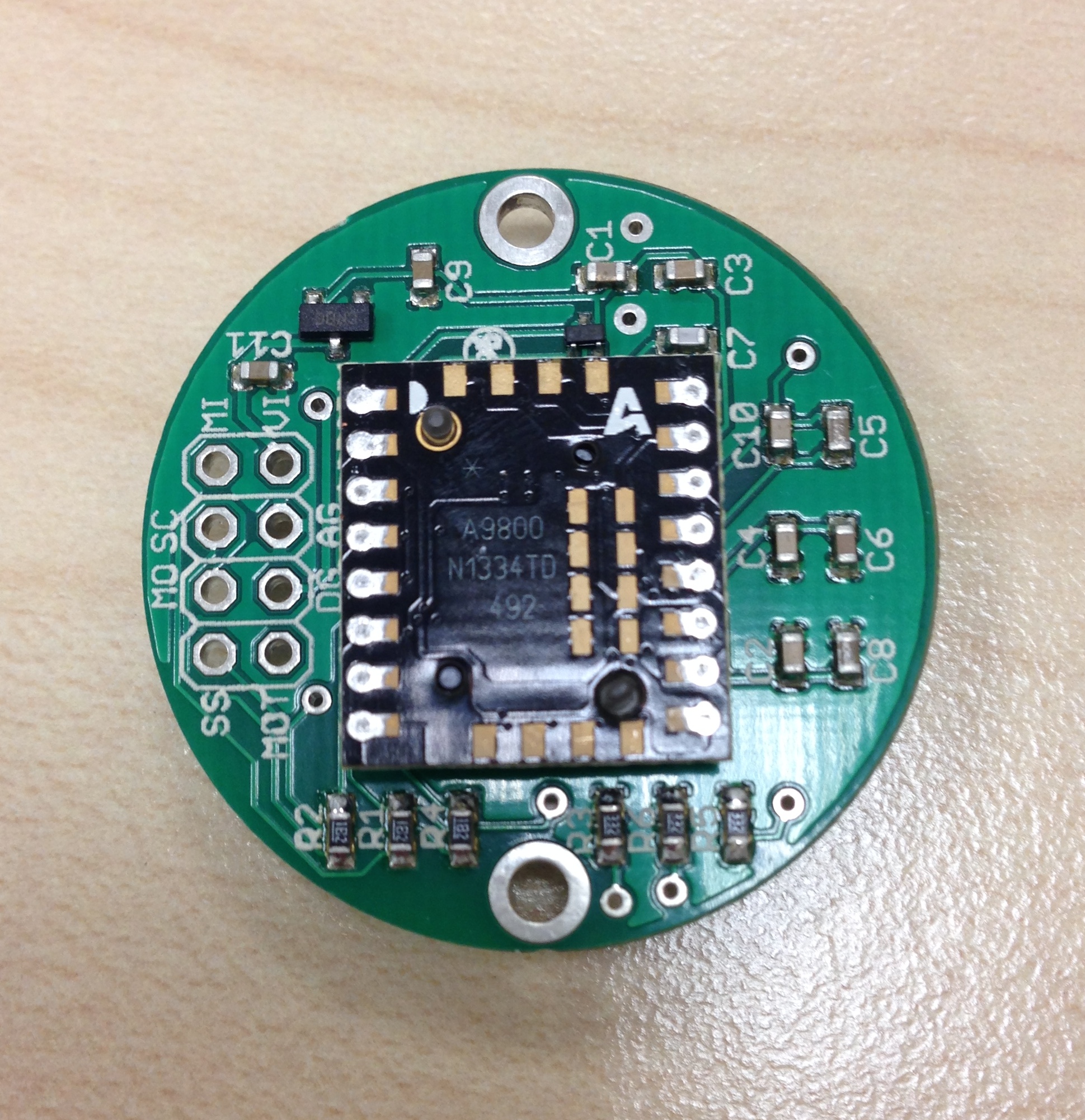}
        \Description{Front view of the optical flow sensor PCB, showing the lens window.}
        \caption{Optical flow sensor front view.}
        \label{subfig:pcb_front}
    \end{subfigure}
    \hfill 
    \begin{subfigure}[b]{0.3\textwidth}
        \centering
        \includegraphics[width=\textwidth]{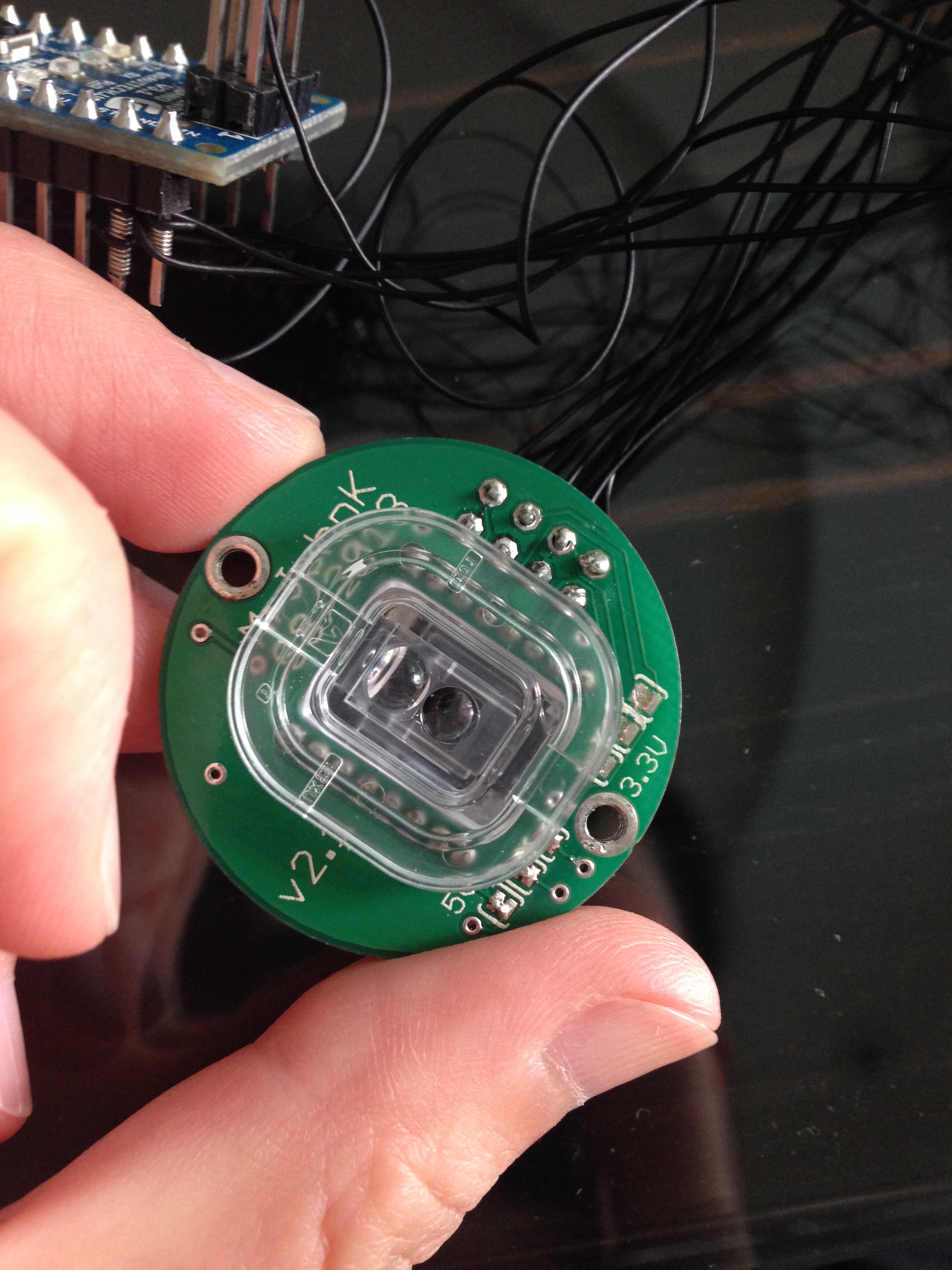}
        \Description{Back view of the optical flow sensor PCB.}
        \caption{Optical flow sensor back view.}
        \label{subfig:pab_back}
    \end{subfigure}
    \caption{Optical Flow Sensor. The ADNS-9800 sensor, equipped with a modified ADNS-6190-002 lens, captures 2D motion data at the base knuckle by analyzing low-resolution images. The front view shows the lens window facing the base knuckle, and the back view shows the pin header that connects the sensor to the Arduino.}
    \label{fig:pcb}
\end{figure}

\begin{figure}[!htbp]
    \centering
    \includegraphics[width=0.5\textwidth]{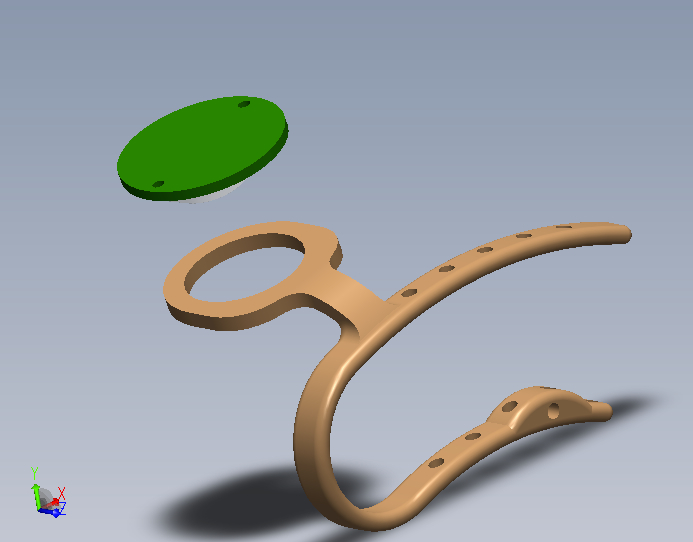}
    \caption{3D-Printed Holding Pin. The customized pin secures the CaRing system to the base knuckle of the index finger, ensuring a minimal distance between the sensor and the finger to improve tracking stability and reduce noise.}
    \Description{A photo of a 3D-printed pin holding the CaRing system on a finger.}
    \label{fig:3d_print_pin}
\end{figure}

Our CaRing prototype uses an ADNS-9800 optical flow sensor with a modified ADNS-6190-002 lens to detect 2D motion at the base knuckle (Figure~\ref{fig:pcb}). This sensor, originally used in optical mice, tracks motion by analyzing low-resolution images it captures. The CaRing system is affixed to the right base knuckle of the index finger with a 3D-printed holding pin (Figure~\ref{fig:3d_print_pin}), shaped for the right hand and holding the sensor at a constant distance from the skin across sessions, reducing noise and keeping the signal comparable across participants.

\subsection{Computer Interfacing}
The optical flow sensor is connected to an Arduino Nano board. A wireless board has limited battery life and suffers from radio interference and dropped connections. To avoid these issues, the board is connected to the host computer, a MacBook Pro running macOS, through a wired USB link, which also powers the board. As a result, every sample arrives at the designated polling rate, and the session length is not limited by a battery. During each session, the host timestamps each sample. Because the researcher's timer runs on the same clock, the recorded signal and the researcher's labels align without correction.

\begin{figure}[!htbp]
    \centering
    \includegraphics[width=0.5\textwidth]{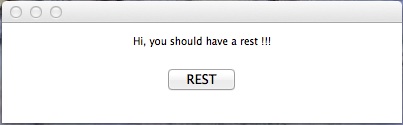}
    \Description{Warning system alert.}
    \caption{Warning system alert. A sample interface pop-up displayed on the user's computer monitor, triggered after fifteen minutes of continuous mouse usage to prompt necessary rest intervals.}
    \label{fig:warning}
\end{figure}

The same host program that receives the sensor data produces the system's output. When CaRing detects a prolonged period of mouse use, the computer beeps and displays a warning message window (Figure~\ref{fig:warning}), prompting the user to take a break. Short breaks taken at regular intervals reduce discomfort from computer work~\cite{mclean2001microbreaks}. Releasing the hand from the mouse relieves the tension that keeps pressure on the median nerve~\cite{cole2006reducing}. In the prototype, the maximum mouse usage period is set to 15 minutes. It is worth emphasizing that the interval is a system parameter, and not a clinical recommendation.

\subsection{Threshold Setting}
\begin{figure}[!htbp]
    \centering
    \includegraphics[width=0.7\textwidth]{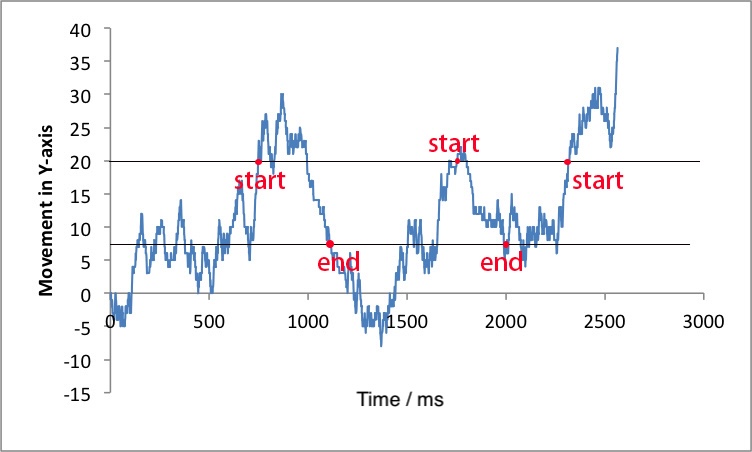}
    \caption{Movement signal with threshold labeling. The graph displays the y-axis motion signal recorded by the CaRing prototype during mouse usage, identifying 'start' and 'end' events via threshold detection after noise smoothing.}
    \Description{A line graph showing the y-axis motion signal with labeled start and end events, as recorded by the CaRing prototype during mouse usage.}
    \label{fig:raw_data_w_label}
\end{figure}

Any hand movement shifts the skin over the base knuckle relative to the sensor window, and the optical flow sensor reports the movement as the difference between two successive frames. Our CaRing system queries the sensor through an Arduino at 12 Hz and adds the reported displacements to a running (x, y) value. The y-axis is aligned with the direction the base knuckle travels when the finger flexes and extends, and the x-axis runs across the finger. Both values are counted from the position the hand holds at the start of the session, not from a fixed point on the desk. When the user starts using the mouse, the system captures a rising edge on the y-axis. When the hand relaxes, the position remains steady, then drops back to baseline (Figure \ref{fig:raw_data_w_label}). The signal is smoothed to remove noise.

\begin{figure}[!htbp]
    \centering
    \includegraphics[width=0.8\linewidth]{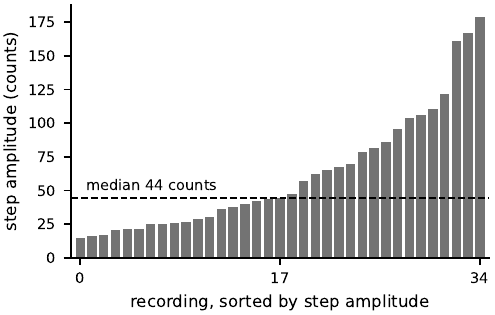}
    \caption{Step amplitude across the 35 recordings, sorted. The step is the height of the sustained y-axis excursion between rest and mouse use, in sensor counts. The spread from 15 to 179 counts is why CaRing sets each session's threshold from that session's own step rather than from a fixed count value.}
    \Description{A bar chart of step amplitude for 35 recordings, sorted from smallest to largest, with a dashed line at the median of 44 counts.}
    \label{fig:step_amp}
\end{figure}

The ADNS-9800 sensor reports relative displacement, not absolute position. Thus, the y-value depends on the size of the user's hand and the pin's position on the knuckle. We recorded 35 sessions with the prototype, totaling 138{,}053 samples, or about 3.2 hours of data at 12 Hz. These recordings are separate from the lab-study sessions in Section~\ref{subsec:verification} and were used only to characterize the signal and choose the parameters, which were fixed before the lab study. For conciseness, we refer to the height of the y-axis signal plateau above the resting level, observed once a person starts using the mouse, as the step. As shown in Figure~\ref{fig:step_amp}, the step ranges from 15 counts to 179 counts across the 35 recordings, and the median is 44 counts. The largest step is 12 times the smallest step. A fixed count threshold therefore cannot serve every user: one high enough for a 179-count step would never be crossed by a 15-count step.

CaRing closes the gap by setting the start and end thresholds as a fixed fraction of each session's step size, rather than as a fixed count. The fraction is 30\%, consistent with the calibration noise analyzed below: the median baseline offset of 3.7 counts is about 25\% of the smallest step of 15 counts, so a lower fraction would place the threshold within the calibration noise of the weakest signal. At the start of a session, the system detects the first clear rise-and-hold in the y-axis signal and measures the plateau height above the session's zero point. The height is taken as the session step. The system detects all later start and end events in the session against 30\% of the value. Thus, the same relative margin applies whether the step is 15 counts or 179 counts.

The step-relative threshold is strongly dependent on how accurately the zero point is measured. Let the zero point be the y-axis signal level when the hand is at rest. However, the reading is rarely exactly zero even when the hand is still. There are two reasons. First, the pin position differs slightly on each knuckle. Second, the sensor output integration drifts. Therefore, we measure the zero point in a 500 ms calibration window at the start of each session. During the window, the user keeps the hand still on the desk. We use the mean y-axis value during the window as the session zero point. The step and the start and end thresholds in Figure~\ref{fig:raw_data_w_label} are measured from the zero point. We choose a 500 ms duration to match the default Run-to-Rest downshift time of the ADNS-9800~\cite{adns9800datasheet}. Thus, the sensor stays at the full frame rate throughout the calibration.

\begin{figure}[!htbp]
    \centering
    \includegraphics[width=0.8\linewidth]{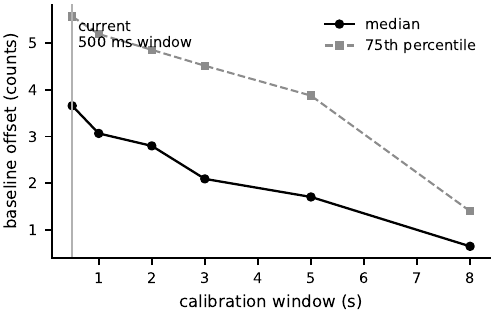}
    \caption{Baseline offset against the length of the calibration window. The offset is the distance from the mean of the first ten seconds of the same recording. A longer window settles closer to the reference, at the cost of a longer wait before a session can start.}
    \Description{A line graph showing median and 75th percentile baseline offset falling as the calibration window grows from 0.5 to 8 seconds.}
    \label{fig:baseline}
\end{figure}

A window longer than 500 ms settles closer to the hand's true resting level. However, it delays the start of every session. Let the baseline offset be the distance from the zero point of a given window to a reference zero point taken over the first 10 seconds of a recording. Figure~\ref{fig:baseline} shows the baseline offset versus the window length, from 0.5 seconds to 8 seconds. The median offset for the 500 ms window is 3.7 counts, compared to 1.7 counts for a 5-second window. The absolute offset is less consequential than its ratio to the threshold margin, which is 30\% of the step. For the median step of 44 counts, the offset is about 28\% of the margin. For the participant with the smallest step of 15 counts, the offset is about 82\% of the margin. Such an offset is enough to move a borderline sample across the threshold. With a 5-second window, this participant's offset drops to about 38\% of the margin, at the cost of a longer wait. A longer window could be used if a user reports missed events.

\section{Results and Analysis}
\label{sec:results}
\subsection{Sensor Placement Test}
We conducted a series of pilot tests to choose the sensor's placement, finger, and orientation.

\subsubsection{Placement and Finger}
In the initial design, we used a wristband with two ADNS-9800 sensors to measure wrist angle. Since the single-threaded Arduino Nano cannot read two sensors at once and the results matched a single sensor, we used one sensor. On the finger, the base knuckle moves more during typing and mouse use than the middle segment, and the pin holds the sensor at a constant distance from the skin there, giving a less noisy signal.

While typing, users are equally likely to use the index or middle finger, but on the mouse the index finger rests on and actuates the left button, so it moves on every click while the middle finger can stay still for long stretches. We therefore placed the sensor on the index finger, where it sees movement during both typing and mouse use.

\subsubsection{Orientation of the Optical Flow Sensor}
\label{subsubsec:orientation}
\begin{figure*}[!htbp]
    \centering
    \includegraphics[width=\linewidth]{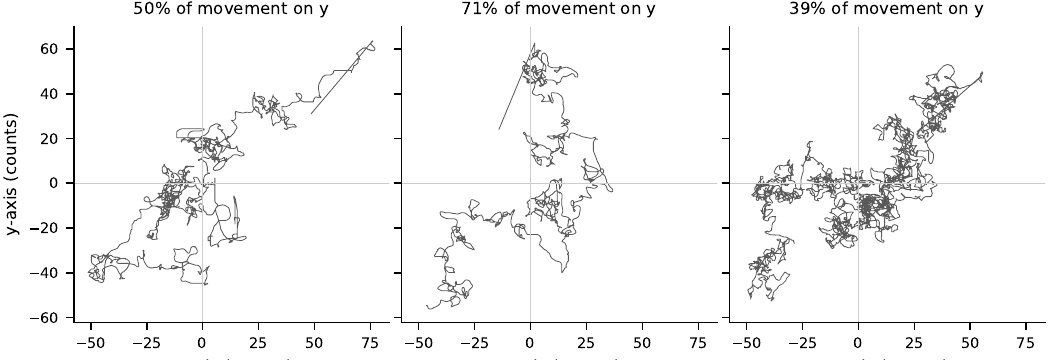}
    \caption{Movement in the sensor plane for the three orientations tested: 0\textdegree, 5\textdegree, and 10\textdegree{} from left to right, matching the panel titles. Each panel plots the smoothed y-axis value against the smoothed x-axis value for one recording. The 5\textdegree{} orientation (middle) captures the largest share of the movement on the y-axis and is the one used for the rest of the paper.}
    \Description{Three scatter panels showing the path of the sensor output in the x and y plane, titled 50\%, 71\%, and 39\% of movement on y for the 0, 5, and 10 degree orientations.}
    \label{fig:orient_data}
\end{figure*}

\begin{figure}[!htbp]
    \centering
    
    \begin{subfigure}[b]{0.45\textwidth}
        \centering
        \includegraphics[width=\textwidth]{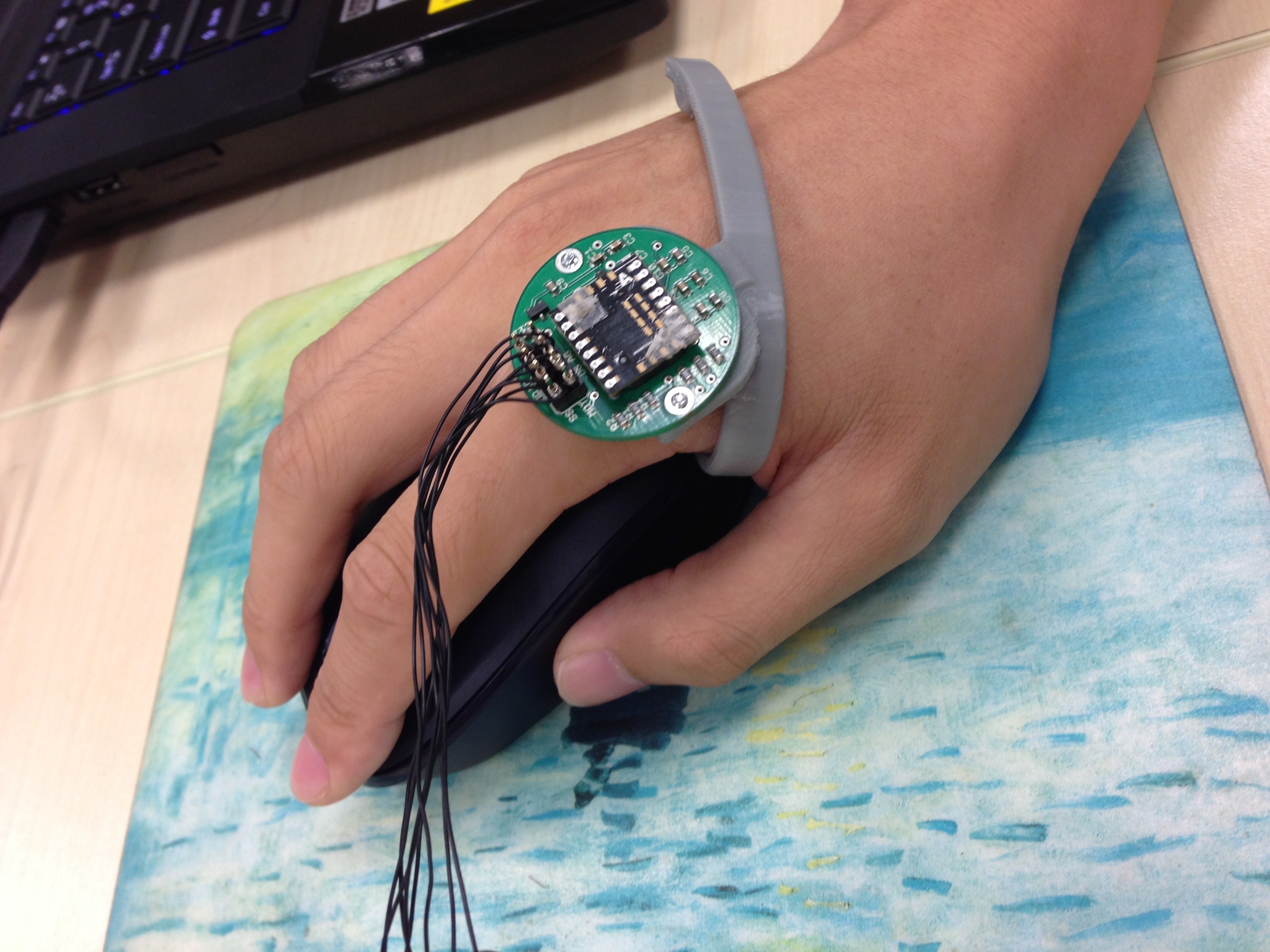}
        \Description{An initial orientation of the optical flow sensor during placement tests.}
        \caption{An initial orientation (0\textdegree) of the optical flow sensor during placement tests.}
        \label{subfig:orentation}
    \end{subfigure}
    \hfill 
    \begin{subfigure}[b]{0.45\textwidth}
        \centering
        \includegraphics[width=\textwidth]{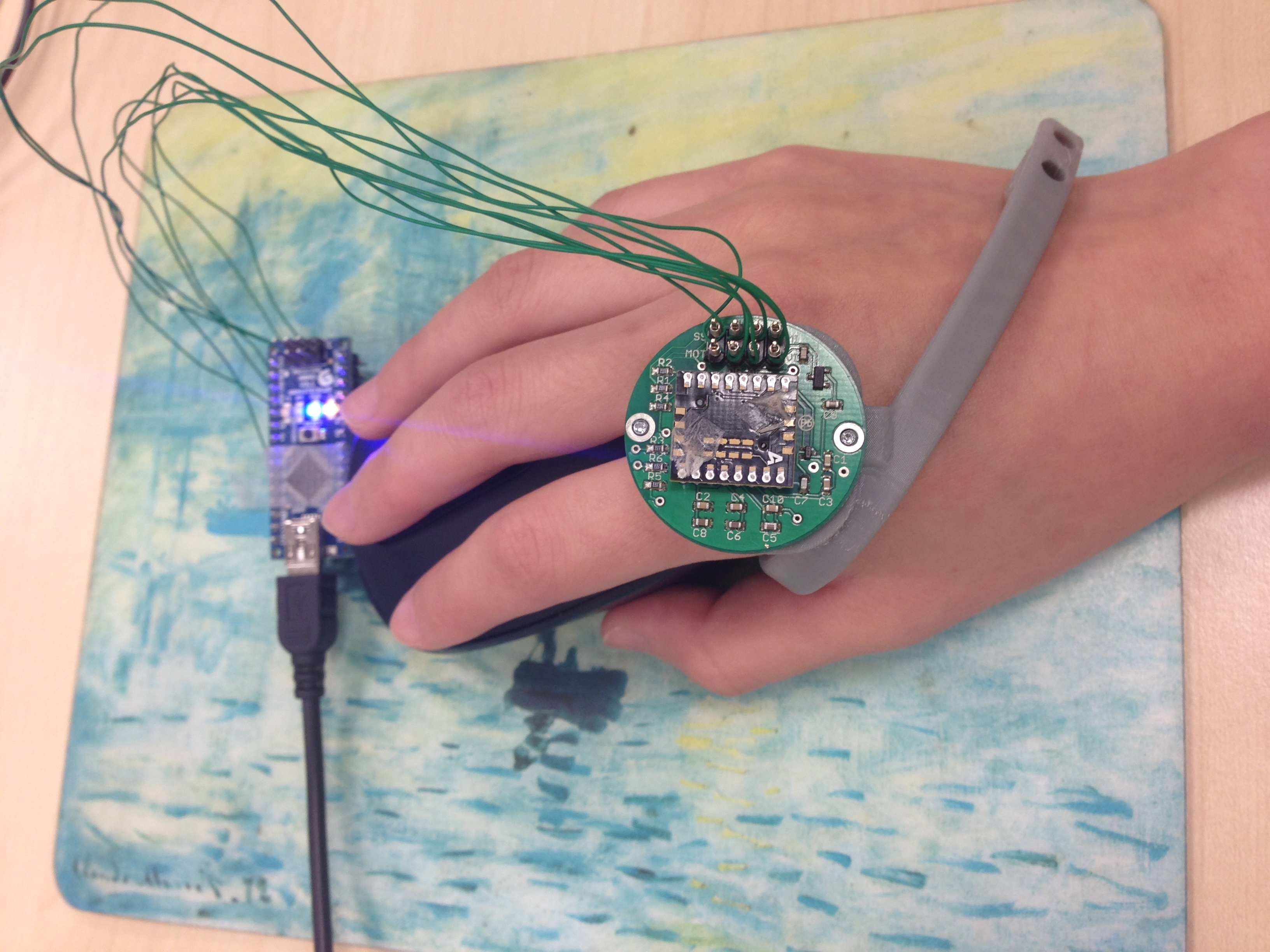}
        \Description{The optimized sensor orientation.}
        \caption{The optimized sensor orientation (5\textdegree), which captured the largest and most repeatable rising edge in our pilot tests.}
        \label{subfig:optimal_orientation}
    \end{subfigure}
    
    \caption{Orientation of the sensor; (b) is the optimized one.}
    \label{fig:orientation}
\end{figure}

Rotating the sensor about the axis normal to the lens window changes which direction of finger motion each axis picks up. Starting from the orientation in Figure~\ref{subfig:orentation}, we recorded the same finger movement at 0\textdegree, 5\textdegree, and 10\textdegree, and computed the share of movement on the y-axis as the range of the smoothed y-axis signal divided by the combined range of the smoothed x- and y-axis signals. As Figure~\ref{fig:orient_data} shows, 0\textdegree, 5\textdegree, and 10\textdegree{} capture 50\%, 71\%, and 39\% of the movement on the y-axis, respectively. We therefore use 5\textdegree{} (Figure~\ref{subfig:optimal_orientation}): it gives the largest, most repeatable rising edge and puts most of the signal on one axis, so the start and end of mouse use can be read from the y-axis alone.

\subsubsection{Axis-specific Movement Analysis}
\begin{figure}[!htbp]
    \centering
    
    \begin{subfigure}[b]{0.45\textwidth}
        \centering
        \includegraphics[width=\textwidth]{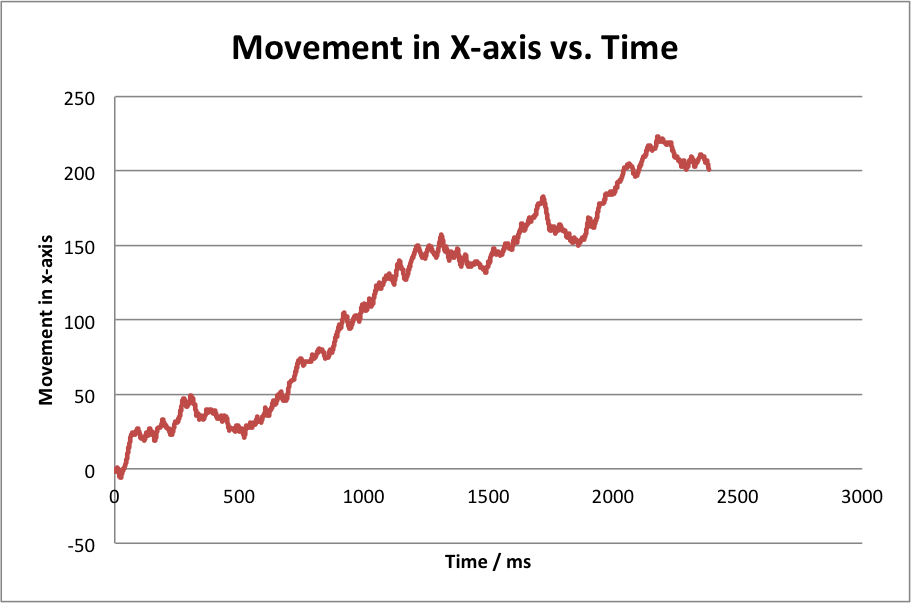}
        \Description{Movement signal along the x-axis over time.}
        \caption{Movement signal along the x-axis over time.}
        \label{fig:x-axis}
    \end{subfigure}
    \hfill 
    \begin{subfigure}[b]{0.45\textwidth}
        \centering
        \includegraphics[width=\textwidth]{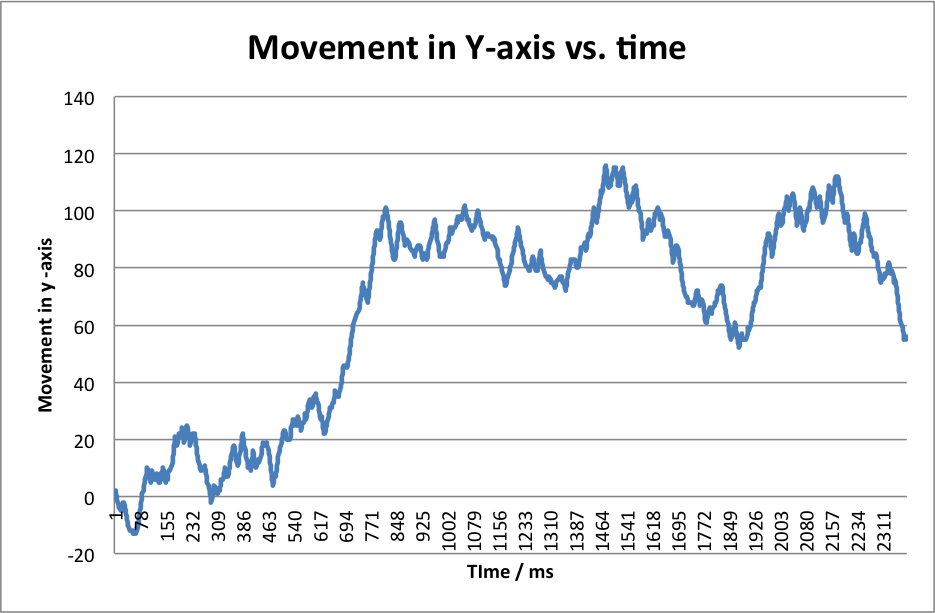}
        \Description{Movement signal along the y-axis over time, with a distinct rising edge at the onset of mouse use.}
        \caption{Movement signal along the y-axis over time, showing the distinct rising edge that characterizes the onset of mouse interaction.}
        \label{fig:y-axis}
    \end{subfigure}
    
    \caption{Axis-specific movement analysis.}
    \label{fig:movement}
\end{figure}

At this orientation, motion appears on both axes when a user starts using the mouse (Figure~\ref{fig:movement}), but only the y-axis carries the rising edge that marks its onset. When the index finger moves onto the mouse button, the base knuckle rises above its resting level, and this one-directional displacement makes the accumulated y-value settle at a higher level and stay there. Motion along the x-axis changes sign as the hand moves the mouse back and forth, so the accumulated value returns to its starting value and no step appears. When the hand leaves the mouse, the base knuckle drops back, and the y-axis returns to the baseline.

\subsection{Verification Test}
\label{subsec:verification}
\begin{figure}[!htbp]
    \centering
    \includegraphics[width=0.8\linewidth]{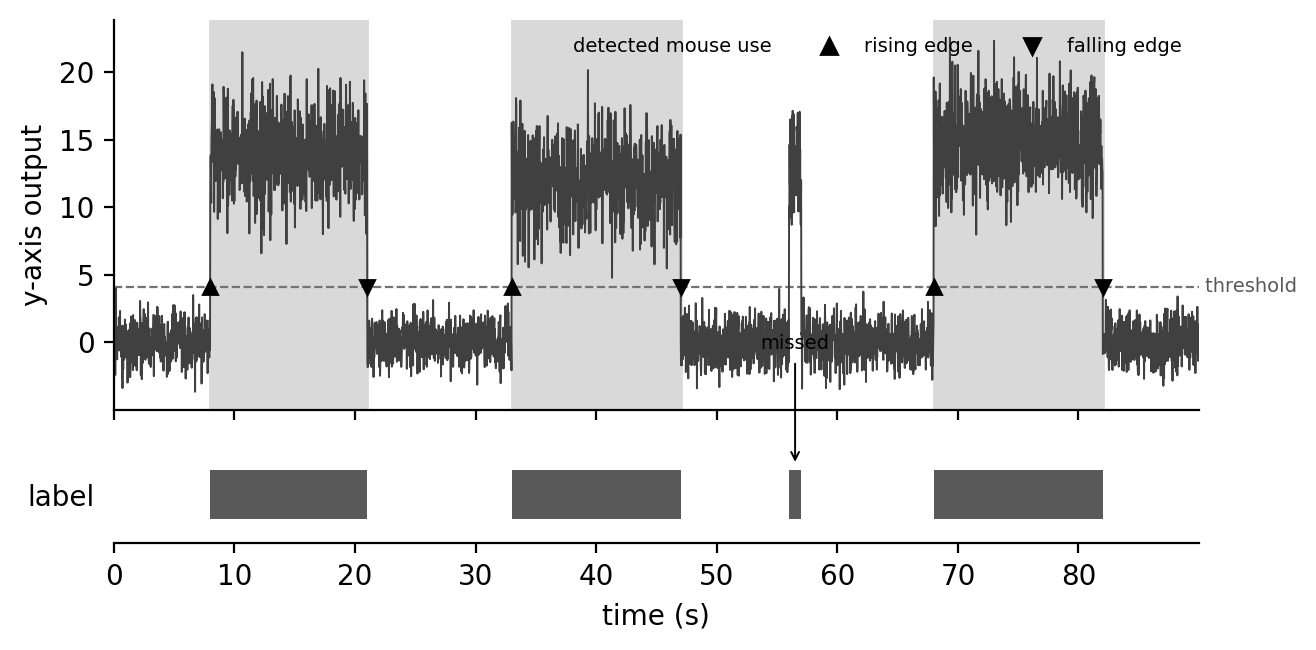}
    \caption{Optical flow signal from one session. The upper trace shows the y-axis output, with the rising and falling edges marked and the detected mouse use segments shaded. The lower bar shows the researcher's labels for the same period.}
    \Description{Optical flow signal from one session, with detected mouse-use segments and the researcher's labels.}
    \label{fig:trace}
\end{figure}

To evaluate CaRing's accuracy in detecting finger movement, we conducted a 30-minute lab study with 10 computer users. The prototype detects the start and end events in real time using the method described in Section~\ref{sec:system}. Figure~\ref{fig:trace} shows one session trace with the detected segments and the researcher's labels.

The participants were students and researchers from the National University of Singapore, aged 20 to 40, with an equal number of female and male participants. All participants were right-handed, which is required by the current prototype. The participants reported about twelve hours of daily computer use on average. Three of the ten participants reported existing pain in the hand or the wrist. Questionnaire items are referenced by number (Q1--Q21); the full questionnaire is given in Appendix~\ref{sec:questionnaire}. Before the session, each participant reported whether their fingers rest on the mouse buttons or whether they release the mouse between actions (Q5). The resting habit is the habit reported in earlier observations of mouse use~\cite{lee2008observed}. We therefore group participants by their answer, keeping the habit separate from the discomfort ratings in Q6 and Q7. Each participant then worked for 30 minutes on self-directed daily tasks, such as research or playing video games, to mimic their daily computer use. These tasks involved both typing and mouse use, so the detection results below include periods of typing.

The participants used the mouse at unplanned moments. Thus, the researcher sat beside the participant with a timer and hand-recorded the beginning and the end of each period of mouse use, as well as the moment the warning appeared. We refer to the record as the researcher's labels, which serve as the ground truth for the system's output.

A detected start or end event is correct if its timestamp is within two seconds of the matching label. The researcher manually marks the labels. Thus, a two-second window allows for reaction time. In the ten sessions, approximately 90\% of the labeled events are correctly detected. The result demonstrates that CaRing reliably recognizes when a user starts and stops using the mouse. Of the remaining events, 3\% are detected late, after the two-second window has passed, and 7\% are not detected at all. The latter case typically occurs because the movement falls under the 30\% margin set for the session (Section~\ref{sec:system}). Such misses are expected to concentrate in sessions with a small step: for the smallest step of 15 counts, the calibration offset alone consumes about 82\% of the margin (Section~\ref{sec:system}).

\begin{figure}[!htbp]
    \centering
    \includegraphics[width=0.8\linewidth]{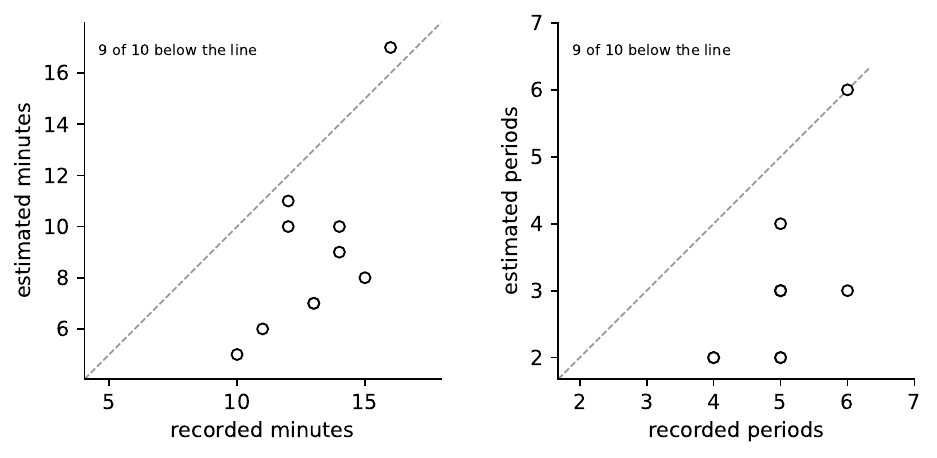}
    \caption{Estimated against recorded mouse usage for each participant. The left panel shows minutes of mouse use, and the right panel shows the number of separate periods of mouse use, each a continuous stretch from a detected start to its matching end. The diagonal line marks a perfect estimate, so points below the line are underestimates.}
    \Description{Estimated against recorded mouse usage of each participant.}
    \label{fig:estimate}
\end{figure}

At the end of each session, participants completed a questionnaire on usage habits, sensor comfort, and perceived work duration (Table~\ref{tab:questionnaire}). Before seeing any sensor output, the participants estimated the duration of mouse use (Q14) and the number of separate periods of use (Q15). The mean estimate was nine minutes, compared with 13 minutes recorded by CaRing, a $1.4\times$ underestimate. Participants estimated three periods on average, compared with five periods recorded. In Figure~\ref{fig:estimate}, we plot each participant's estimate against the recorded value. In each panel, nine of ten points fall below the line of a perfect estimate. 
Participants underestimated both measures. On average, they recalled 
$4.0$ fewer minutes than the record showed, with a $95\%$ CI of $2.2$ to $5.8$ minutes, and a Wilcoxon signed-rank test gives $W = 1.5$, $p = .006$. All nine participants who misjudged the number of periods guessed too few, and a sign test gives $p = .004$.
After reviewing the sensor log, seven of ten participants agreed that the reported periods matched those they remembered (Q17). Thus, the participants trust the log once it is presented to them, even though their unaided estimate was low.

In Q6 and Q7, the participants rated the discomfort of resting their fingers on the mouse buttons and of releasing the mouse between actions. Seven of ten participants rated the resting posture higher, with a mean of 3.4 on a five-point scale, compared to 2.1. In Q8, no participant reported any new pain during the session. Thus, the ratings reflect the general discomfort associated with the habit, rather than an effect induced by the lab session. It is worth emphasizing that only ten self-assessed ratings were collected. We therefore treat the pattern as motivation for a longer field study rather than as evidence that a single habit causes more pain.

Six of ten participants found the sensor comfortable to wear. Eight participants would wear the sensor if it were wire-free. Comfort over a full working day was not tested; the responses suggest that removing the wire would address much of the remaining hesitation.

Nine of ten participants noticed the warning message (Q18), and seven participants changed the hand posture or took a break (Q19). Six participants considered the fifteen-minute interval to be about right (Q20). However, the responses are self-reported and not observed. The result shows that the warning reaches users and that most users act on it once. It does not show that the warning changes the posture over a full working day.

\begin{table}[!htbp]
    \centering
    \begin{tabular}{@{}lc@{}}
        \toprule
        \textbf{Questionnaire item} & \textbf{Participants} \\
        \midrule
        \multicolumn{2}{@{}l}{\textit{Pain and habit (Q6, Q7)}} \\
        More pain with the fingers on the mouse   & 7 \\
        \addlinespace
        \multicolumn{2}{@{}l}{\textit{Comfort and interference (Q9--Q13)}} \\
        Found the sensor comfortable to wear      & 6 \\
        Ring interfered with using the mouse      & 4 \\
        Ring interfered with typing               & 3 \\
        Would wear the ring without a wire        & 8 \\
        \addlinespace
        \multicolumn{2}{@{}l}{\textit{Sensor log accuracy (Q17)}} \\
        Reported periods matched what they remembered & 7 \\
        \addlinespace
        \multicolumn{2}{@{}l}{\textit{The warning (Q18, Q19)}} \\
        Noticed the warning message               & 9 \\
        Changed hand posture after the warning    & 4 \\
        Took a break after the warning            & 3 \\
        Did nothing after the warning             & 3 \\
        \addlinespace
        \multicolumn{2}{@{}l}{\textit{The interval (Q20)}} \\
        Fifteen minute interval too short         & 3 \\
        Fifteen minute interval about right       & 6 \\
        Fifteen minute interval too long          & 1 \\
        \bottomrule
    \end{tabular}
    \caption{Questionnaire results from the thirty-minute lab study. Each value is the number of participants out of ten who gave the response. For the five-point items, a rating of four or five counts as agreement.}
    \label{tab:questionnaire}
\end{table}

Taken together, the lab study and questionnaire suggest that a finger-worn sensor can detect periods of mouse use, and that participants tend to underestimate this usage without it. Whether the resulting warnings reduce computer-related injuries in daily life is a question our thirty-minute lab study cannot answer, and we return to it in Section~\ref{subsec:future_work}.
 
\section{Discussion}
\label{sec:discussion}
This section discusses two uses of CaRing and the limitations that motivate future work.

\subsection{Toward Always-Available Monitoring}
The first is an always-available monitoring device that tracks daily computer usage habits. It requires only that the user wear the ring on the base knuckle of the index finger while using the computer. The system detects the movement of the finger and warns the user when a period of use reaches the set interval, fifteen minutes in the current prototype. The prototype warns on duration alone; telling a harmful posture from a safe one requires the independent posture ground truth discussed in Section~\ref{subsec:future_work}. A key advantage of CaRing, discussed in Section~\ref{sec:related}, is that it senses the finger directly rather than elapsed time or wrist angle, and our verification test shows this signal alone detects the start and end of mouse use with about 90\% accuracy. We have not compared CaRing against these approaches in the same sessions, and such a comparison is left to future work. For the user, the warning marks when a break is due, and taking that break is exactly the behavior that prevention advice asks for~\cite{mayoclinic_carpaltunnel_2026}.

\subsection{Data Collection for Medical Research}
The second use is data collection for medical research. CaRing records when users start and stop using the mouse, which produces a usage log that self-reports cannot provide. When a patient reports pain in the hand or forearm, the doctor relies on the patient's account of how long and how often they used the mouse. Our lab study suggests this account runs low by about a third. This log replaces that account with a recorded duration and period count, which can feed risk metrics such as the Hand Activity Level~\cite{radwin2023}. No medical researcher or doctor participated in the design or evaluation of the current prototype, so we have not tested the log's clinical value.

\subsection{Future Work}
\label{subsec:future_work}
The current prototype and its evaluation have three limitations. We discuss each limitation along with the corresponding future work.

Firstly, the evaluation is short and small in scale. Each session lasts thirty minutes, and a single researcher records the labels by hand. In future work, a longer field study will be conducted with more participants and a second rater.

Second, we evaluate only mouse detection. The evaluation does not assess whether a posture is harmful or whether the warning corrects the posture. In addition, CaRing is not compared with the software-sensing systems in Section~\ref{subsec:software_sense} in the same sessions. In a longer study, the posture should be scored independently, and CaRing should be run alongside these systems for a direct comparison.

Third, the prototype connects to the host through a wired USB link, and the holding pin is printed in a single size for the right hand. In future work, the wired link can be replaced by a Bluetooth Low Energy link. Adjustable and mirrored pins can be designed to fit a wider range of knuckle sizes and the left hand.
 
\section{Conclusion}
\label{sec:conclusion}
Carpal Tunnel Syndrome can cause permanent damage to the median nerve. This paper established foundational work in detecting mouse use from finger movement. Accurate detection of mouse use is critical to Carpal Tunnel Syndrome prevention. As a first step, we introduced CaRing, a ring worn on the index finger. To the authors' best knowledge, CaRing is the first system for Carpal Tunnel Syndrome prevention that detects mouse use by sensing finger movement. Thus, neither a specially designed mouse nor a camera is required. Furthermore, the reminder is issued after fifteen minutes of measured mouse use instead of elapsed time. We then propose a detection method designed for two distinct signal features. Firstly, the amplitude differs by up to a factor of 12 across users. Secondly, the reading is rarely zero when the hand is at rest. Because of these two features, we measure a zero point at the start of each session and set the detection threshold to 30\% of the step amplitude. With this formulation, no per-user parameter needs manual tuning. We set the design parameters using 35 recordings. We validated CaRing's efficacy in a separate lab study with ten users. CaRing detects approximately 90\% of start and end events within two seconds of the researcher's label. The participants underestimated their mouse use by about 4 minutes in a thirty-minute session. The CaRing log can help users schedule breaks and help researchers study hand injury.

\section{Acknowledgments}
Keio-NUS CUTE (Connective Ubiquitous Technology for Embodiments) Center at the National University of Singapore funded the work. We thank all the volunteers, publications support, and staff who wrote and provided helpful comments.
 
\bibliographystyle{ACM-Reference-Format}
\bibliography{ref}

@article{tosti2012acute,
  title={Acute carpal tunnel syndrome},
  author={Tosti, R. and Ilyas, A. M.},
  journal={Orthop Clin North Am},
  volume={43},
  number={4},
  pages={459--465},
  year={2012},
  month={Oct}
}

@misc{mayoclinic_carpaltunnel_2026,
  author       = {{Mayo Clinic Staff}},
  title        = {Carpal Tunnel Syndrome: Symptoms and Causes},
  howpublished = {\url{https://mayoclinic.org}},
  year         = {2026},
  note         = {Accessed: 2026-07-15}
}

@article{cole2006reducing,
  title={Reducing musculoskeletal burden through ergonomic program implementation in a large newspaper},
  author={Cole, Donald C. and Hogg-Johnson, Sheilah and Manno, Michael and Ibrahim, Selahadin and Wells, Richard P. and Ferrier, Sue E. and {Worksite Upper Extremity Research Group}},
  journal={International Archives of Occupational and Environmental Health},
  volume={80},
  number={2},
  pages={98--108},
  year={2006}
}

@article{lee2008observed,
  title={Observed finger behavior during computer mouse use},
  author={Lee, D. and McLoone, H. and Jindrich, D.},
  journal={Applied Ergonomics},
  volume={39},
  pages={107--113},
  year={2008}
}

@misc{rsiguard,
  author = {{Cority}}, 
  title = {RSIGuard},
  year = {2026},
  url = {http://www.rsiguard.com},
  lastaccessed = {July 11, 2026}
}

@inproceedings{jaimes2005sit,
author = {Jaimes, Alejandro},
title = {Sit straight (and tell me what I did today): a human posture alarm and activity summarization system},
year = {2005},
isbn = {1595932461},
publisher = {Association for Computing Machinery},
address = {New York, NY, USA},
url = {https://doi.org/10.1145/1099083.1099087},
doi = {10.1145/1099083.1099087},
booktitle = {Proceedings of the 2nd ACM Workshop on Continuous Archival and Retrieval of Personal Experiences},
pages = {23–34},
numpages = {12},
location = {Hilton, Singapore},
series = {CARPE '05}
}

@inproceedings{leung2012limber,
author = {Leung, Ken and Reilly, Derek and Hartman, Kate and Stein, Suzanne and Westecott, Emma},
title = {Limber: DIY wearables for reducing risk of office injury},
year = {2012},
isbn = {9781450311748},
publisher = {Association for Computing Machinery},
address = {New York, NY, USA},
url = {https://doi.org/10.1145/2148131.2148150},
doi = {10.1145/2148131.2148150},
booktitle = {Proceedings of the Sixth International Conference on Tangible, Embedded and Embodied Interaction},
pages = {85–86},
numpages = {2},
location = {Kingston, Ontario, Canada},
series = {TEI '12}
}

@inproceedings{tobias2013exploring,
author = {Sonne, Tobias and Gr\o{}nb\ae{}k, Kaj},
title = {Exploring new potentials in preventing unhealthy computer habits},
year = {2013},
isbn = {9781450319522},
publisher = {Association for Computing Machinery},
address = {New York, NY, USA},
url = {https://doi.org/10.1145/2468356.2468442},
doi = {10.1145/2468356.2468442},
booktitle = {CHI '13 Extended Abstracts on Human Factors in Computing Systems},
pages = {487–492},
numpages = {6},
location = {Paris, France},
series = {CHI EA '13}
}

@article{bhaskar2004ergonomic,
author = {Gupta, Bhaskar},
title = {Ergonomic soft mouse and armrest mouse pad},
year = {2004},
issue_date = {September 2004},
publisher = {Association for Computing Machinery},
address = {New York, NY, USA},
volume = {2004},
number = {September},
url = {https://doi.org/10.1145/1029383.1029385},
doi = {10.1145/1029383.1029385},
journal = {Ubiquity},
month = sep,
pages = {2},
numpages = {1}
}

@article{watanabe2021,
  title={The Accuracy of a Screening System for Carpal Tunnel Syndrome Using Hand Drawing},
  author={Watanabe, Takuro and Koyama, Takafumi and Yamada, Eriku and Nimura, Akimoto and Fujita, Koji and Sugiura, Yuta},
  journal={Journal of Clinical Medicine}, volume={10}, number={19}, pages={4437}, year={2021},
  doi={10.3390/jcm10194437}, publisher={MDPI}}

@article{moschetti2016,
  title={Recognition of Daily Gestures with Wearable Inertial Rings and Bracelets},
  author={Moschetti, Alessandra and Fiorini, Laura and Esposito, Dario and Dario, Paolo and Cavallo, Filippo},
  journal={Sensors}, volume={16}, number={8}, pages={1341}, year={2016},
  doi={10.3390/s16081341}, publisher={MDPI}}

@article{radwin2023,
  title={Comparison of the Observer, Single-Frame Video and Computer Vision Hand Activity Levels},
  author={Radwin, Robert G. and Hu, Yu Hen and Akkas, Oguz and Bao, Stephen and Harris-Adamson, Carisa and Lin, Jia-Hua and Meyers, Alysha R. and Rempel, David},
  journal={Ergonomics}, volume={66}, number={8}, pages={1132--1141}, year={2023},
  doi={10.1080/00140139.2022.2136407}, publisher={Taylor \& Francis}}

@article{mennella2022,
  title={Characteristics and Applications of Technology-Aided Hand Functional Assessment: A Systematic Review},
  author={Mennella, Ciro and Alloisio, Susanna and Novellino, Antonio and Viti, Federica},
  journal={Sensors}, volume={22}, number={1}, pages={199}, year={2022},
  doi={10.3390/s22010199}, publisher={MDPI}}

@manual{adns9800datasheet,
  title        = {ADNS-9800 LaserStream Gaming Sensor Data Sheet},
  organization = {PixArt Imaging},
  number       = {ADNS-9800},
  year         = {2013},
  url          = {https://pixart.com}
}

@article{mclean2001microbreaks,
  author    = {McLean, L.},
  title     = {Computer terminal work and the benefit of microbreaks},
  journal   = {Applied Ergonomics},
  volume    = {32},
  number    = {3},
  pages     = {225--237},
  year      = {2001},
  publisher = {Elsevier}
}

@inproceedings{mateevitsi2014healthbar,
author = {Mateevitsi, Victor and Reda, Khairi and Leigh, Jason and Johnson, Andrew},
title = {The health bar: a persuasive ambient display to improve the office worker's well being},
year = {2014},
isbn = {9781450327619},
publisher = {Association for Computing Machinery},
address = {New York, NY, USA},
url = {https://doi.org/10.1145/2582051.2582072},
doi = {10.1145/2582051.2582072},
booktitle = {Proceedings of the 5th Augmented Human International Conference},
articleno = {21},
numpages = {2},
location = {Kobe, Japan},
series = {AH '14}
}

@inproceedings{gu2025mocount,
author = {Gu, Ruocheng and Jia, Sen and Ma, Yule and Zhong, Jinqin and Hwang, Jenq-Neng and Li, Lei},
title = {MoCount: Motion-Based Repetitive Action Counting},
year = {2025},
isbn = {9798400720352},
publisher = {Association for Computing Machinery},
address = {New York, NY, USA},
url = {https://doi.org/10.1145/3746027.3755857},
doi = {10.1145/3746027.3755857},
booktitle = {Proceedings of the 33rd ACM International Conference on Multimedia},
pages = {9026–9034},
numpages = {9},
location = {Dublin, Ireland},
series = {MM '25}
}

@article{JIA2026115077,
title = {A unified multimodal framework for human behavior understanding via motion and language alignment},
journal = {Engineering Applications of Artificial Intelligence},
volume = {178},
pages = {115077},
year = {2026},
issn = {0952-1976},
doi = {https://doi.org/10.1016/j.engappai.2026.115077},
url = {https://www.sciencedirect.com/science/article/pii/S0952197626013606},
author = {Sen Jia and Hao Zhang and Wen Zhao},
}

@inproceedings{he2025abacus,
  title={Simulating Human Cognition in Abacus Gesture Learning: An ACT-R with Vision/Motor and PyIBL Approach},
  author={He, Lingyun and Tehranchi, Farnaz},
  booktitle={Proceedings of the 23rd International Conference on Cognitive Modeling (ICCM)},
  pages={61--67}, year={2025}}

@inproceedings{yuan2022opticalflow,
  title={Optical Flow Training under Limited Label Budget via Active Learning},
  author={Yuan, Shuai and Sun, Xian and Kim, Hannah and Yu, Shuzhi and Tomasi, Carlo},
  booktitle={European Conference on Computer Vision (ECCV)},
  pages={410--427}, year={2022}, publisher={Springer}}

@inproceedings{ni2024timeseries,
  author={Ni, Haowei and Meng, Shuchen and Geng, Xieming and Li, Panfeng and Li, Zhuoying and Chen, Xupeng and Wang, Xiaotong and Zhang, Shiyao},
  booktitle={2024 6th International Conference on Electronic Engineering and Informatics (EEI)}, 
  title={Time Series Modeling for Heart Rate Prediction: From ARIMA to Transformers}, 
  year={2024},
  pages={584-589},
  doi={10.1109/EEI63073.2024.10695966},
  publisher={IEEE}}

@ARTICLE{hu2025infocons,
  author={Hu, Wang and Mohsenian-Rad, Hamed and Farrell, Jay A.},
  journal={IEEE Transactions on Vehicular Technology}, 
  title={{Optimization-Based Outlier Accommodation Using Information Constraints for CAV State Estimation in Urban Environments}}, 
  year={2025},
  volume={},
  number={},
  pages={1-15},
  doi={10.1109/TVT.2025.3569207}}

@misc{han2026earlyearlyenoughdesigndependent,
      title={How Early Is Early Enough? Design-Dependent Observation-Window Sufficiency in Subscription Churn Prediction}, 
      author={Xiao Han and Yao Xiao and Chenyu Wu and Tongchen Zhang},
      year={2026},
      eprint={2607.00473},
      archivePrefix={arXiv},
      primaryClass={cs.LG},
      url={https://arxiv.org/abs/2607.00473}, 
}

@misc{zhang2026liralocalcrosslayerinformation,
title={LIRA: Local Cross-Layer Information Routing for Vision-Language-Action Decoding},
author={Zhewei Zhang and Puyue Wang and Guanren Qiao and Yijie Weng and Jiawei Hu and Guo Li and Lujia Wang and Junyan Wang and Tao Gu and Hongliang Lu and Guiliang Liu and Hong Jia and Xinhu Zheng},
year={2026},
eprint={2608.07596},
archivePrefix={arXiv},
primaryClass={cs.RO},
url={arxiv.org/abs/2608.07596},
}

@article{liang2026oc,
  title={OC-Distill: Ontology-aware Contrastive Learning with Cross-Modal Distillation for ICU Risk Prediction},
  author={Liang, Zhongyuan and Jo, Junhyung and Lee, Hyang-Jung and Kim, Sang Kyu and Chen, Irene Y},
  journal={arXiv preprint arXiv:2604.16878},
  year={2026}
}

@article{ginsberg2022learning,
  title={A learning based hypothesis test for harmful covariate shift},
  author={Ginsberg, Tom and Liang, Zhongyuan and Krishnan, Rahul G},
  journal={arXiv preprint arXiv:2212.02742},
  year={2022}
}

@article{zhou2024reducing,
  title={Reducing manual labeling requirements and improved retinal ganglion cell identification in 3D AO-OCT volumes using semi-supervised learning},
  author={Zhou, Mengxi and Zhang, Yue and Karimi Monsefi, Amin and Choi, Stacey S and Doble, Nathan and Parthasarathy, Srinivasan and Ramnath, Rajiv},
  journal={Biomedical Optics Express},
  volume={15},
  number={8},
  pages={4540--4556},
  year={2024},
  publisher={Optica Publishing Group}
}

@misc{xu2026chainawareencodingmicroservicetrace,
      title={Chain-Aware Encoding for Microservice Trace Anomaly Detection},
      author={Yiliu Xu and Ziwei Hong and Zhongheng Yang and Xinjin Li and Yu Ma},
      year={2026},
      eprint={2607.10156},
      archivePrefix={arXiv},
      primaryClass={cs.SE},
      url={https://arxiv.org/abs/2607.10156},
}

@inproceedings{chang2025pdunet,
  title     = {Research on PDU-Net Lung Nodule Segmentation Algorithm Based on Path Aggregation and Dual Attention},
  author    = {ChiaHua Chang and M. Fu and X. Chen and S. Feng and M. Zhang and X. Zhou and J. J. Chuang and J. Pan and Z. Liu},
  booktitle = {2025 4th International Conference on Image Processing, Computer Vision and Machine Learning (ICICML)},
  address   = {Chongqing, China},
  year      = {2025},
  publisher = {IEEE},
  isbn      = {979-8-3315-6581-7},
  doi       = {10.1109/ICICML67980.2025.11333821},
  note      = {IEEE conference paper, first author (1 of 9); added to IEEE Xplore Jan 19, 2026; lung nodule segmentation, U-Net, path aggregation network, dual attention, LIDC-IDRI, LUNA16, medical imaging AI}
}

@inproceedings{
zeng2026does,
title={Does reflective {LLM} feedback survive a wet-lab-scale label budget?},
author={Haoxuan Zeng and XIN LUO},
booktitle={NeurIPS 2026 Agentic AI for Biological Discovery Workshop},
year={2026},
url={https://openreview.net/forum?id=ezGV5BIXVJ}
}
 
\appendix
 
\section{Questionnaire}
\label{sec:questionnaire}
 
Participants answered $Q1$ to $Q5$ before the session began. $Q6$ to $Q13$ and $Q18$ to $Q21$ were answered at the end of each session. $Q14$ to $Q16$ were also answered at the end of the session, before the participant saw any sensor output, and $Q17$ was answered afterward. Items marked (1--5) use a five-point scale, where 1 is the lowest and 5 is the highest.

 \subsection*{Background}
\begin{enumerate}
  \item How many hours a day do you use a computer?
  \item Which hand do you use for the mouse?
    \item Do you have, or have you had, pain, numbness, or tingling in your hand, wrist, or forearm? If yes, for how long?
 
      \item Have you been diagnosed with Carpal Tunnel Syndrome or another hand or wrist condition?
  \item When you are not clicking, do you rest your fingers on the mouse buttons, or do you release the mouse?
  \end{enumerate}
 
\subsection*{Pain and Habit}
 
\begin{enumerate}[resume]
  \item Rate the pain or discomfort in your hand and forearm when you keep your fingers on the mouse. (1--5)
  \item Rate the pain or discomfort when you release the mouse between actions. (1--5)
  \item Did you feel any pain or discomfort during the thirty-minute session?
\end{enumerate}
 
\subsection*{Comfort and Interference}
\begin{enumerate}[resume]
  \item Was the ring comfortable to wear? (1--5)
  \item Did the ring interfere with using the mouse? (1--5)
  \item Did the ring interfere with typing? (1--5)
  \item Would you wear the ring for a full working day?
  \item Would you wear it if it had no wire?
\end{enumerate}
 
\subsection*{Usage Estimate}
 
\begin{enumerate}[resume]
  \item Without looking at the sensor output, how many minutes do you think you used the mouse during the session?
  \item How many separate periods of mouse use do you think you had?
  \item How confident are you in those two estimates? (1--5)
  \item Now that you have seen the sensor log, did the reported periods match what you remember? (1--5)
\end{enumerate}
 
\subsection*{The Warning}
 
\begin{enumerate}[resume]
  \item Did you notice the warning when it appeared?
  \item What did you do after the warning appeared? Nothing, changed hand posture, took a break, or something else.
  \item Was the time too short, about right, or too long?
  \item Would you want the system to suggest what to do, or only tell you to rest?
\end{enumerate}
 
\end{document}